\documentclass[aps,pra,twocolumn,showpacs]{revtex4}

\newcommand{\Jeff}{J_\mathrm{eff}}
\newcommand{\polya}{\Xi^{\ast}}

\usepackage{graphicx}

\begin{document}

\title{Finding zeros of the Riemann zeta
function by periodic driving of cold atoms}

\author{C.E.~Creffield}
\affiliation{Departamento de F\'isica de Materiales, Universidad
Complutense de Madrid, E-28040, Madrid, Spain}

\author{G.~Sierra}
\affiliation{Instituto de F\'isica Te\'orica, UAM-CSIC, E-28049, Madrid, Spain}

\date{\today}


\begin{abstract}
The Riemann hypothesis, which states that the non-trivial
zeros of the Riemann zeta function all lie on a certain line
in the complex plane, is one of the most important unresolved
problems in mathematics. 
We propose here a new approach to finding a physical system to study
the Riemann zeros, which in contrast to previous examples,
is based on applying a time-periodic driving field.
This driving allows us to tune the quasienergies of the system (the analogue
of the eigenenergies for static systems), so that they are
directly governed by the zeta function. We further show
by numerical simulations that this allows the Riemann zeros to
be measured in currently accessible cold atom experiments.
\end{abstract}

\maketitle

\section{Introduction}
The Riemann hypothesis states that the non-trivial zeros of the
Riemann function, $\zeta(s)$, have the form $s_n = 1/2 \pm i E_n$,
where the $E_n$ are all real.
A fascinating approach to treating the Riemann problem is
to consider the $E_n$ to be eigenvalues of a self-adjoint operator,
since if such an operator could be identified the $E_n$ would necessarily be
real. This idea, known as the P\'olya and Hilbert conjecture, is supported by
numerous evidence (see \cite{S10,SH11} for reviews), notably
that the zeros appear to closely follow the GUE statistics \cite{gue1,gue,B86,K99}
of random matrix theory and quantum chaos. This unexpected connection
between number theory and physics has inspired many suggestions for
finding physical implementations of the Riemann function, 
such as evaluating the Fourier transform of a suitable
wave using opto-mechanical means \cite{van_der_pol} or
far-field diffraction \cite{berry_antenna}, or by measuring entanglement
in quantum systems \cite{schleich}.

In this work we propose a very different
approach, in which a time-periodic driving potential is used
to modify the dynamics of a quantum system. As the system is
periodically-driven its dynamics is not described by energy
eigenvalues, but by a generalization of these quantities termed
``quasienergies''. Our central result is the construction of a
driving field for which the corresponding quasienergy spectrum is given
by the Riemann $\Xi$ function \cite{P26,E74}, or by a smoothed
version of this function, $\polya$, introduced by P\'olya. This last has
the appealing feature that the appropriate driving field can be expressed in
a simple closed form.
In both cases the zeros of the $\Xi$ function correspond to degeneracies,
or crossings, of the quasienergies. These
are of particular physical significance because they correspond
to the phenomenon known as ``coherent destruction of tunneling'' (CDT)
in which the dynamics of the system is frozen \cite{cdt}.
As an example we show how the zeros can be
seen directly in cold atom experiments by measuring the expansion
rate of a condensate held in a driven optical lattice.
This technique thus represents a new and powerful way of
finding a physical realization of the Riemann function.

\section{Method} 
We begin by considering a standard two-level system,
driven by a time-periodic function $f(t) = f(t+T)$
\begin{equation}
H(t) = -J \ \sigma_x + f(t) / 2 \ \sigma_z  \ ,
\label{2level}
\end{equation}
where $J$ is the tunneling between the two levels.
As $f(t)$ is time-periodic, the natural framework
to treat the problem is given by Floquet theory \cite{review}.
In this approach one seeks the eigensystem of
the Floquet operator
\begin{equation}
{\cal H}(t) = H(t) - i \partial_t \ ,
\label{floquet}
\end{equation}
where we have set $\hbar = 1$. Henceforth we shall
also measure all energies (and frequencies) in units of $J$.
The eigenstates of ${\cal H}(t)$ are $T$-periodic
functions called Floquet states, and their associated
eigenvalues, which play an analogous role to energy eigenvalues
for the case of a static Hamiltonian, are called quasienergies.
The Floquet states provide a complete basis, and expanding
the wavefunction in these states provides similar advantages
to the normal procedure of expressing a state in energy
eigenstates in the undriven case.

In general it is difficult to obtain analytical
expressions for the Floquet states.
In the strong-driving limit, however, when the frequency
$\omega = 2 \pi /T$ is the dominant energy scale,
it is possible to make an expansion
by first solving just for the time-dependent component of $H(t)$,
and applying the static part as perturbation \cite{cec_prb}.
In this way one obtains a perturbative series in orders of $J$.
Truncating at first-order gives the simple result
$H_\mathrm{eff} = - \Jeff  \ \sigma_x$, where
the effective tunneling $\Jeff$ is given by
\begin{equation}
\Jeff / J = \frac{1}{T} \int_0^T dt \ e^{-i F(t)} \ ,
\label{integral_eff}
\end{equation}
and $F(t) = \int_0^t dt' f(t')$. The quasienergies are
given simply by the eigenvalues of $H_\mathrm{eff}$,
namely $\epsilon_\pm = \pm | \Jeff |$.

\begin{figure}
\begin{center}
\includegraphics[width=0.45\textwidth,clip=true]{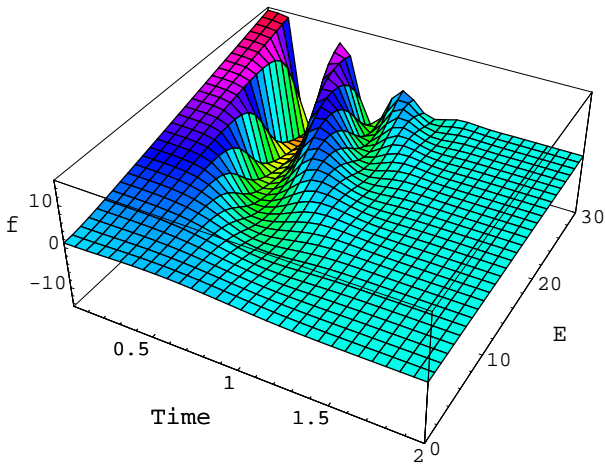}
\end{center}
\caption{Surface plot of the time-dependent driving function $f(t)$
for the smoothed function $\polya(E)$ (Eq.\ref{final}).
For small values of $E$ the function is quite featureless, but progressively
increases in amplitude and develops more oscillations
as $E$ increases. The corresponding driving function for $\Xi(E)$ behaves
similarly (see Fig.\ref{surface_plot}.)}
\label{drive_function}
\end{figure}

From Eq.\ref{integral_eff} it is thus straightforward to calculate
the behavior of the quasienergies (or equivalently, of the
effective tunneling) for a given driving potential $f(t)$.
For the case $f(t) = K \cos \omega t$, for example, this
yields the well-known Bessel function renormalization \cite{shirley} of 
tunneling
$\Jeff = J {\cal J}_0(K/\omega)$. At zeros of the Bessel function,
$K / \omega = 2.404, \ 5.520 \dots \ $, the effective tunneling
vanishes, producing CDT. This effect has been measured experimentally
\cite{pisa,oberthaler,eckardt} in the dynamics of driven
ultracold atoms.

We need, however, to solve the {\em inverse problem}; to find an $f(t)$
that produces a given behavior of the quasienergies.
We shall first explain our technique using
P\'olya's function, $\polya(E)$, as this gives a convenient closed-form
for the solution, and then go on to consider the more
complicated case of the true Riemann $\Xi$ function.

\subsection{Driving function for P\'olya's function}
P\'olya's function is given by \cite{footnote_factor}
\begin{equation}
\polya (E) = 4 \pi^2 \left( K_{a + i E/2}(x) +
K_{a - i E/2}(x) \right) \ ,
\label{zeta}
\end{equation}
where $K_\beta(t)$ is the modified $K$-Bessel function, $x=2 \pi$ 
and $a = 9/4$.
This is a smoothed version of the Riemann $\Xi$ function
\begin{equation}
\Xi(E) = \frac{1}{2} s (s-1) \Gamma \left( \frac{s}{2} \right) \pi^{- s/2} \zeta(s),
\qquad s = 1/2 + i E
\label{rie}
\end{equation}
whose zeros coincide with the non-trivial zeros of $\zeta(s)$.
Although as Titchmarsh pointed out \cite{T86} P\'olya's
zeta function cannot truly be regarded as an approximation
to $\Xi(E)$ in the most obvious sense, they do 
share many properties. Most importantly $\polya(E)$ 
has the same average distribution of zeros \cite{E74,T86}, 
following the smooth term of the Riemann-Mangoldt formula,
and so nonetheless represents an interesting application of our method.
P\'olya further proved that the zeros of $\polya(E)$ are real for
any value of the constant $a$.
The spectrum of the $xp$-type \cite{sierra}  and Dirac Hamiltonian \cite{S14}
is given, for example, by the zeros of (\ref{zeta})
with $a=1/2$.

The modified
Bessel function can be conveniently expressed as the integral identity
\begin{equation}
K_\beta \left( x \right) = \int_0^\infty dt \cosh \left( \beta t \right)
e^{-x \cosh t} \ ,
\label{bessel}
\end{equation}
and thus
\begin{eqnarray}
\polya(E) &=& 8 \pi^2 \Re \left[ K_{a + i E/2}(2 \pi) \right] \nonumber \\
&=& 8 \pi^2 \int_0^\infty dt \cosh \left( a t \right)
e^{-2 \pi \cosh t} \cos \left( E t / 2	\right) .
\label{bessel_useful}
\end{eqnarray}

We are aiming to obtain the result
$\Re \left[ \Jeff(E) \right] \propto \polya(E)$.
Combining Eqs.\ref{integral_eff} and \ref{bessel_useful}
reveals that this requires $F(t)$ to obey the relation
\begin{equation}
\int_0^T  dt \ \cos F(t) =
\int_0^\infty dt \ \alpha \cosh \left( a t \right)
e^{-2 \pi \cosh t} \cos \left( E t / 2 \right) \ ,
\label{requirement}
\end{equation}
where $\alpha$ is an arbitrary constant.
As the integrand on the right-hand side
decays rapidly with $t$, we can replace the upper
limit of integration with $T$, as long as we take
$T$ to be sufficiently large.
In this case we can then write \cite{footnote}
\begin{equation}
F(t) = \cos^{-1} \left[ \alpha	\cosh (a t) e^{-2 \pi \cosh t}
\cos  (E t / 2) \right] \ .
\label{define_F}
\end{equation}
The value of $\alpha$ is now fixed by noting
that we require  $F(0) = 0$, and thus $\alpha = e^{2 \pi}$.
The driving field is then given by $f(t) = \partial_t F(t)$, yielding
the final result
\begin{equation}
f(t) = - \frac{
\phi(t)
\left(a \tanh a t - 2 \pi \sinh t - E / 2 \ \tan (E t / 2) \right)}
{\sqrt{1 - \phi(t)^2}}
\label{final}
\end{equation}
where $\phi(t) =  \cosh (a t) e^{2 \pi (1 - \cosh t)} \cos (E t / 2)$.

\begin{figure}
\begin{center}
\includegraphics[width=0.45\textwidth,clip=true]{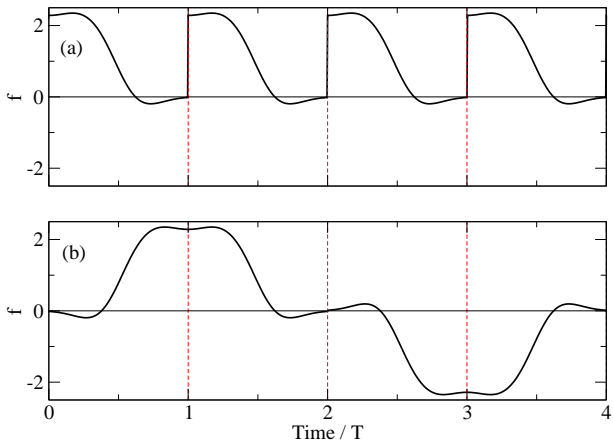}
\end{center}
\caption{(a) The driving potential, $f(t)$, is extended to be
a periodic function by periodically repeating its behavior
over the interval $0 \leq t < T$. This produces, however,
a function with rather poor performance.
(b) Construction of a more efficient driving potential. Four copies
of $f(t)$ are joined together, following some reflection
transformations, to create a single continuous function,
resembling two localized pulses of opposite sign.
The vertical dashed lines indicate where the segments are joined.
This function is continuous, has a
time-average of zero, and $f(t) = - f(t + T_0/2)$
where $T_0 = 4 T$ is the total period of the signal,
satisfying the parity requirement given in the text.}
\label{driving}
\end{figure}

In Fig.\ref{drive_function} we show the form of $f(t)$
as $E$ is varied. Little structure is visible for $t > 1.5$.
The reason is the rapid decrease of $\phi(t)$ with $t$, arising from
the exponential term $e^{ - 2 \pi \cosh t}$.
Indeed $|\phi(t)| < 1.3 \times 10^{-6}, \;  \forall E$ and $t >2$.
Accordingly we now set the period of the
driving, and thus the cut-off in the integration in Eq.\ref{requirement},
to be $T=\pi/2$. We shall use this value throughout the rest of the
paper. In Fig.\ref{driving}a we show
the full periodic driving field, obtained
by periodically repeating cycles of $f(t)$.
Although this form of $f(t)$ indeed satisfies (\ref{requirement}), the
driving can be made more effective by imposing a further set
of conditions on it: {\it i)} to avoid heating in the cold atom model 
(see Section \ref{BEC}), the
average of $f(t)$ over one period should vanish,
{\it ii)} discontinuities should be avoided, and
{\it iii)} for the quasienergy crossings to be well-defined,
the Floquet states must be from different parity classes.
If condition {\it (iii)} is not fulfilled, the von Neumann-Wigner theorem
implies that the quasienergies cannot cross as $E$ is varied, and will
instead form a broad avoided crossing. As the system is
periodically-driven, the appropriate generalized parity operator involves
both inversion and time-translation
${\cal P}:x \to -x, \ t \to t + T/2$.
These three conditions can be satisfied by joining
four copies of the fundamental waveform (\ref{final}) as
shown in Fig.\ref{driving}b, to create what amounts to
two pulses of opposite sign, with a total period of $T_0 = 4 T$.

\subsection{Driving function for $\Xi(E)$}
\begin{figure}[b]
\begin{center}
\includegraphics[width=0.45\textwidth,clip=true]{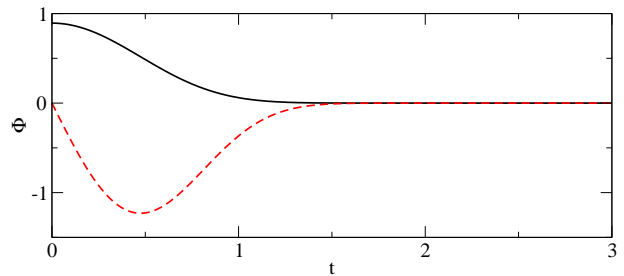}
\end{center}
\caption{Behaviour of $\Phi(t)$ and its derivative.
Solid black line: $\Phi(t)$ has a global maximum at $t=0$, for
which $\Phi(0) \simeq 0.89339$, and decays rapidly as $t$ increases.
Dashed red line: the derivative, $\Phi'(t) = \partial_t \Phi(t)$,
evaluated from Eq.\ref{def_deriv}. The series expansions for both $\Phi(t)$
and $\Phi'(t)$ were truncated at 30 terms.}
\label{Phi}
\end{figure}

We now turn to the more important case of the Riemann function itself.
This can be written
\begin{equation}
\Xi(E) = \int_0^{\infty} dt \ \Phi(t) \cos \left(E t /2 \right) \ ,
\label{true_zeta}
\end{equation}
where $\Phi(t) = 2 \pi e^{5 t /4} \sum_{n=1}^{\infty}
\left( 2 \pi e^t n^2 - 3 \right) n^2 e^{- \pi n^2 e^t}$,
as defined in \cite{P26}.
We now follow the same procedure as before, seeking
a driving potential such that $\Re \left[ \Jeff(E) \right] \propto \Xi(E)$.
This requires finding the solution of the equation
\begin{equation}
\int_0^T  dt \ \cos F(t) = \alpha
\int_0^\infty dt \ \Phi(t) \cos \left( E t / 2 \right) \ ,
\label{require_true}
\end{equation}
where $\alpha$ is a constant.
We show the behavior of $\Phi(t)$ in Fig.\ref{Phi}.
Since $\Phi(t)$ decreases rapidly with $t$,
being fitted reasonably well by a simple Gaussian function 
\cite{berry_antenna}, we can substitute
the upper limit of integration on the right-hand side of Eq.\ref{require_true}
to be $T$, as long as $T$ is sufficiently large for the integral to
converge. As before, we take $T = \pi/2$. We can then write
\begin{equation}
F(t) = \cos^{-1} \left[ \alpha \Phi(t)
\cos  (E t / 2) \right] \ .
\label{define_F_true}
\end{equation}
Imposing the boundary condition $F(0)=0$ requires $\alpha$
to take the value $\alpha = 1 / \Phi(0)$. As $\Phi(0)$ is the
global maximum of the function, this guarantees that $|\alpha \Phi(t)| \leq 1$
for all values of $t$, and thus $F(t)$ is real and well-defined.

Having obtained $F(t)$ it is now straightforward to calculate
$f(t) = \partial_t F(t)$. This yields the final result
\begin{equation}
f(t) = - \frac{\left[
\Phi'(t) \cos \left( E t / 2 \right) -
\left( E / 2 \right) \Phi(t) \sin \left( E t / 2 \right) \right]}
{\sqrt{
\Phi^2(0) - \left[ \Phi(t) \cos \left( E t / 2 \right) \right]^2
}} \ .
\label{def_f}
\end{equation}
The derivative of $\Phi(t)$ can be evaluated by differentiating
its series expansion term by term, to give
\begin{equation}
\Phi'(t) = \frac{\pi e^{5 t / 4}}{2}
\sum_{n=1}^{\infty} \left(
30 e^t n^2 \pi - 8 e^{2 t} n^4 \pi^2  - 15 \right)
n^2 e^{- \pi n^2 e^t} \ ,
\label{def_deriv}
\end{equation}
which we also show in Fig.\ref{Phi}.

\begin{figure}
\begin{center}
\includegraphics[width=0.45\textwidth,clip=true]{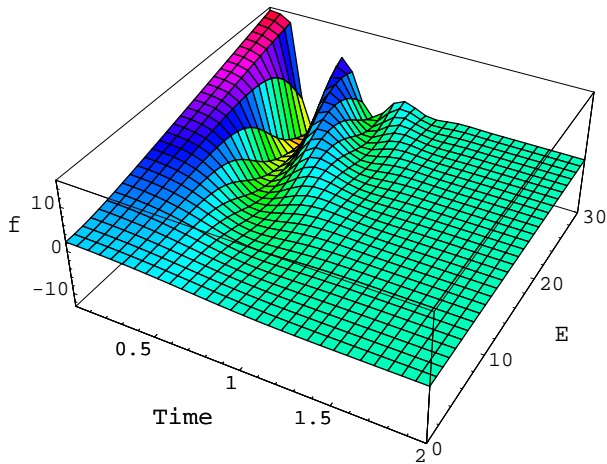}
\end{center}
\caption{Surface of the driving function $f(t)$ (Eq.\ref{def_f}),
which produces a quasienergy spectrum proportional to the
Riemann $\Xi$ function. The behaviour resembles strongly
the driving function for the smoothed Riemann function, shown in
Fig.\ref{drive_function}.}
\label{surface_plot}

\end{figure}

We show the full behavior of $f(t)$ as a function of time and the
parameter $E$ in Fig.\ref{surface_plot}.
Despite the complicated form of (\ref{def_f}),
this plot shows a striking similarity to Fig.\ref{drive_function},
in which the $f(t)$ giving rise to $\polya(E)$ is shown.
This should not be unexpected as $\polya(E)$ is simply a smoothed
version of the Riemann function, and consequently the main features
of the driving functions must be the same.

\begin{figure}
\begin{center}
\includegraphics[width=0.45\textwidth,clip=true]{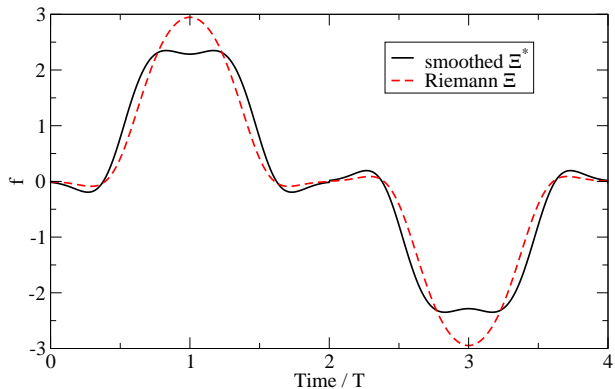}
\end{center}
\caption{Although the driving functions, $f(t)$, that produce $\polya(E)$ and
$\Xi(E)$ look superficially very similar, they differ in details. We show here
the driving functions for $E=4$ for the two Riemann functions we consider.}
\label{compare}
\end{figure}

Just as for the case of the smoothed Riemann function, a more effective
driving potential is obtained by joining four copies of $f(t)$,
to produce a continuous function with a definite parity and
zero time average. In Fig.\ref{compare}
we show the form of the driving function we obtain in this way,
and compare it with the driving function that produces $\polya(E)$.
Examining the two curves in detail, reveals that
although the functions share the same general form
(as seen in the similarity of the two surface plots
Fig.\ref{drive_function} and Fig.\ref{surface_plot}),
there are nonetheless significant small differences. These minor
differences lead to the different location of zeros for
$\Xi(E)$ and $\polya(E)$, demonstrating the well-known
analytical sensitivity of the zeta function.

\section{Results}

\subsection{High frequency limit}
The expression for the effective tunneling, $\Jeff$, that we use in this
work comes from a first-order perturbation theory calculation of
the driven system. In this perturbation theory, the ``small parameter''
is $J / \omega$, and so this result is only valid in the high-frequency
limit $\omega \gg J$. This would require the driving period, $T = 2 \pi/\omega$,
to be small; however this contradicts the requirement that $T$ be as large
as possible so that the integrals (\ref{requirement}) and (\ref{require_true}) 
are well-converged.
To be able to satisfy both these requirement, we therefore
scale the driving as $f(t) \to \Omega f(\Omega t)$, where
$\Omega > 1$,  while keeping $T_0 = 2 \pi$ constant. 
Note that this is not a trivial rescaling of time, as its effect is
to make the pulses shown in Fig.\ref{driving}b narrower
and taller, while keeping their spacing constant at $T$.

\begin{figure}
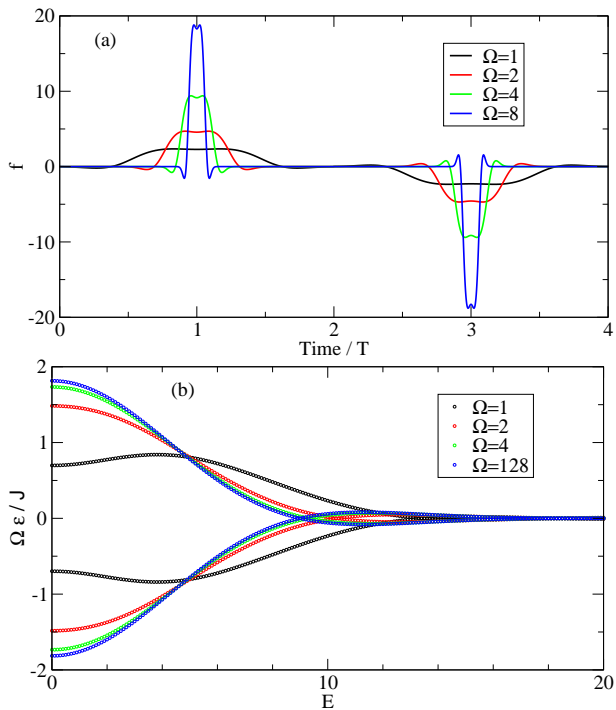

\begin{center}
\includegraphics[width=0.45\textwidth,clip=true]{fig6a}
\includegraphics[width=0.45\textwidth,clip=true]{fig6b}
\end{center}
\caption{(a) Change of the driving function as $\Omega$ is
increased. The pulses become progressively narrower and taller,
while their spacing remains the same.
(b) As $\Omega$ increases, the quasienergy spectrum asymptotically
approaches that of the infinite frequency limit, and the quasienergy
crossings move towards the zeros of the Riemann function.
Here we show the quasienergies for a system driven to reproduce
the smoothed Riemann function, $\polya(E)$. Note how the first crossing
asymptotically approaches the first zero of $\polya(E)$ at $E=8.993$
as $\Omega$ increases.}
\label{pulse}
\end{figure}

We show the effect of increasing $\Omega$ on the pulse-shape in Fig.\ref{pulse}a.
It can be clearly seen that the pulses become progressively more
localized and of higher amplitude.
As the pulses become shorter and more intense, the quasienergy spectrum
of the system approaches that of the high-frequency limit, as we show
in Fig.\ref{pulse}b. In particular we can see that the
locations of the quasienergy crossings converge toward the
zeros of the Riemann function. This evolution of the quasienergy spectrum
as the driving frequency increases is a general feature
of periodically-driven systems, including for example,
sinusoidal, squarewave, and triangular driving \cite{cec_prb}.

In principle $\Omega$ should be made as large as possible, to ensure that the
system is well within the high-frequency regime. 
We can see from Fig.\ref{pulse}b that
good precision is obtained for $\Omega > 4$, and the results
we present below use $\Omega=128$. Increasing $\Omega$ beyond
this value was found to introduce instabilities
in the numerical integration of the system's time-evolution.

\subsection{Quasienergies}
\begin{figure}
\begin{center}
\includegraphics[width=0.45\textwidth,clip=true]{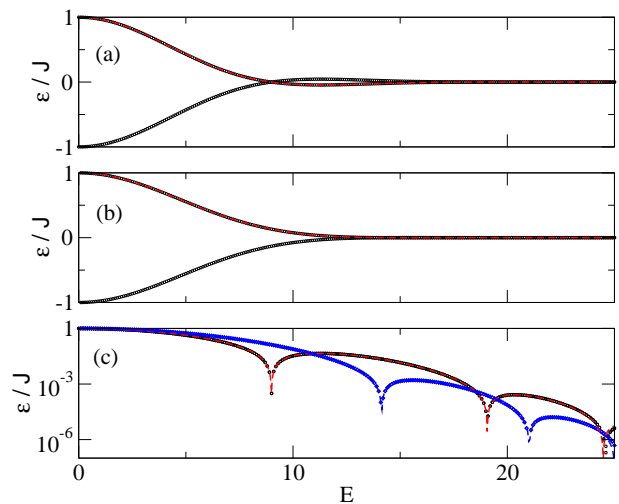}
\end{center}
\caption{(a) Black symbols show the quasienergies of the
two-level system driven by the potential $f(t)$ given
by Eq.\ref{final}, the dashed (red) line is the
smoothed Riemann function $\polya(E)$.
The quasienergies are normalized with respect to the
quasienergies at $E=0$. The agreement between the
two is excellent.
(b) As above, but for the true Riemann $\Xi$ function,
with the driving potential given by Eq.\ref{def_f}.
Again, the results agree perfectly.
(c) Comparison on a logarithmic scale. The first three zeros
of $\polya(E)$, corresponding to degeneracies of
the quasienergies, are clearly visible as the downward-pointing cusps
(black circles, red dashed line),
as are the first two zeros of $\Xi(E)$
(blue diamonds, blue dashed line) at $E\simeq 14.1347$ and $21.0220$.
The quasienergies reproduce the behavior of the
$\Xi$ functions to a high degree of accuracy.}
\label{quasi}
\end{figure}

In Fig.\ref{quasi}a we compare the quasienergies, obtained by
the direct integration of the equation of motion (\ref{2level})
under the driving potential given in Eq.\ref{final},
with the exact behavior of $\polya(E)$. The agreement is
seen to be excellent. Similarly in Fig.\ref{quasi}b
we compare the Riemann $\Xi$ function (\ref{true_zeta}) with
the quasienergies resulting from driving the system
with the potential given by Eq.\ref{def_f},
and again see essentially perfect agreement.
As the $\Xi$ functions decay roughly exponentially with
$E$ \cite{sierra}, we show in
Fig.\ref{quasi}c the same data plotted logarithmically.
The cusps visible in this plot
correspond to zeros of the $\Xi$ functions, and thus to
crossings of the quasienergies at which $\Jeff$ vanishes.
We see that
the quasienergies accurately reproduce the
behavior of the $\Xi$ functions over at least six orders of magnitude,
although eventually precision effects do lead to deviations
at large values of $E$.

\subsection{\label{BEC} Measuring the effective tunneling}
A way of directly measuring $\Jeff$ in experiment is to observe
the expansion of a gas of cold atoms \cite{expand,eckardt,pisa}.
If the atoms are held in an optical lattice potential they
can be described well by a tight-binding model
\begin{equation}
H_0 = -J \sum_{\langle i, j \rangle}
\left( a_i^\dagger a_j + \mathrm{H. c.} \right) +
\sum_j V(r_j) n_j \ ,
\end{equation}
where $i$ labels the lattice site,
$\langle i, j \rangle$ are nearest-neighbors,
and $V(r)$ is a trap potential. By ``shaking'' the optical
lattice \cite{goldman}
it is possible to introduce a time-periodic driving potential
$H(t) = H_0 + f(t) \sum_j r_j n_j$ which generalizes Eq.\ref{2level}
from a two-level model describing two sites to the case of
$N$ lattice sites. For a parabolic trap potential, the initial
state of the system will be Gaussian. If the trap potential
is then released, this Gaussian wavepacket will undergo free expansion
at a rate governed by $| \Jeff |$ \cite{expand}.

\begin{figure}
\begin{center}
\includegraphics[width=0.45\textwidth,clip=true]{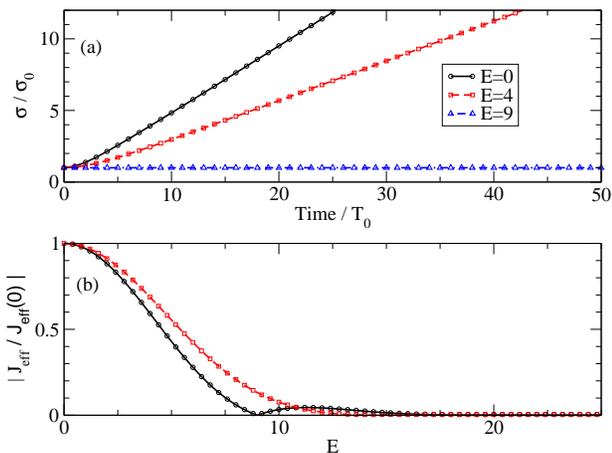}
\end{center}
\caption{(a) Expansion of a Gaussian wavepacket in an optical lattice
with $N=200$ sites under the periodic driving potential $f(t)$ (\ref{final}),
such that $\Jeff \propto \polya(E)$.
For $E=0$ the wavepacket
spreads quickly, but for $E=4$ the wavepacket spreads less rapidly, indicating
that $\Jeff$ is smaller. For $E=9$ the wavepacket hardly expands at
all, indicating that CDT is occurring.
(b) The solid black curve shows $|\polya(E)|$, and the black
circles the values of $| \Jeff |$ measured from the expansion curves.
The agreement between the two is excellent. Similarly the dotted red line
shows $|\Xi(E)|$, and the red squares the measured values of $|\Jeff|$
for a system driven by the potential (\ref{def_f}).
Again the agreement is perfect.}
\label{expand}
\end{figure}

In Fig.\ref{expand}a we show the spread of the wavepacket with time,
$\sigma(t) = \sqrt{\langle x^2 \rangle - \langle x \rangle^2}$,
under the periodic driving corresponding to $\polya(E)$.
For $E=0$ the wavepacket expands rapidly, and soon enters
the ballistic regime in
which $\sigma \propto t$. For $E=4$ the expansion is slower since
$\Jeff$ is smaller, and for $E=9$ the
wavepacket barely expands at all, indicating that this value of
$E$ is close to a zero of $\polya(E)$.
In principle one would expect only partial destruction
of tunneling, even when $E$ is tuned exactly to a zero of $\polya$,
due to the presence of longer-ranged hopping elements.
Such an effect is seen in sinusoidally-driven systems \cite{eckardt}
where the band strongly narrows but does not collapse totally.
For the case of the drivings we consider, the longer-ranged
hoppings are suppressed considerably more than for the sinusoidal
case, however, meaning their effect is essentially negligible.

These estimations can be made quantitative. In Fig.\ref{expand}b we
show the values of $|\Jeff|$ obtained by measuring the spread
of the wavepacket after an expansion time of $50 T_0$ and fitting
it to the result $|\Jeff| \propto \sqrt{\sigma(t)^2 - \sigma_0^2}$,
where $\sigma_0$ is the initial width of the condensate.
Both of the driving potentials show excellent agreement
with the exact $\Xi$ functions, demonstrating
that measuring the expansion of a trapped cold atom system is
a viable method to experimentally determine the zeros of the Riemann
functions.

\section{Conclusions}
We propose a novel approach for a physical realization
of the Riemann zeros.
The main idea is to use a time-dependent
driving potential to modify the dynamics of a quantum system,
so that its quasienergy spectrum mimics the desired Riemann function.
We provide a systematic scheme for calculating
the appropriate potential, and suggest a physical realization
using driven cold atoms.
Although the observation of high Riemann zeros is hindered by the rapid
decay of the $\Xi$ function, this could be compensated in experiment by
increasing the tunneling strength (which depends exponentially on the optical
lattice depth), or by using a form of the zeta function which
decays more slowly, such  as $\xi(s)/s$ \cite{van_der_pol,berry_antenna}.
An intriguing possibility for future study would be to apply
this technique to more general trigonometric integrals
of the function $\Phi(t)$, which could thereby lead to the measurement
of a new bound on the de Bruijn-Newman constant, the value of which
is related to the Riemann hypothesis \cite{odlyzko}.

\bigskip

\acknowledgments
The authors thank Michael Berry for stimulating discussions.
CEC was supported by the Spanish MINECO through
Grant Nos. FIS2010-21372 and FIS2013-41716-P, and GS by Grant No.
FIS2012-33642, QUITEMAD, and the Severo Ochoa Programme
under grant SEV-2012-0249.


\begin{thebibliography}{99}
\bibitem{footnote}
{Since we employ the relation $F(t) = \cos^{-1} \phi(t)$, we clearly
require $|\phi(t)| \leq 1$ for all $t$. It can be shown that a necessary
and sufficient condition for this is $a^2 < x$, where $x$ and $a$ parameterize
the modified Bessel function (\ref{zeta}). For the case
we are interested in, $a=9/4$ and $x = 2 \pi $  (e.g. $2.25 < 2.50$) , the relation
$a < \sqrt{x}$ is indeed satisfied.}
\end{thebibliography}
\end{document}